\documentclass{aa}

\usepackage{graphics}
\begin{document}

  % \thesaurus{03
   %           (13.07.1;  % Gamma-Ray: Bursts
    %           )}
   \title{Limits on the early afterglow phase\\
   of gamma-ray burst sources from TAROT-1}

   \subtitle{}

   \titlerunning{Early afterglow of GRBs}

   \author{M. Bo\"er\inst{1} \and J.L. Atteia\inst{1} \and M. Bringer\inst{1}
\and B. Gendre\inst{1} \and A.
   Klotz\inst{1} \and R. Malina\inst{2} \and J.A. de Freitas Pacheco\inst{3}
   \and H. Pedersen\inst{4}
          }

   \offprints{M. Bo\"er}
   \mail{M. Bo\"er}

   \institute{Centre d'Etude Spatiale des Rayonnements (CNRS/UPS), BP 4346,
   31028 Toulouse Cedex 4, France,\\
              email: Michel.Boer@cesr.fr
\and Laboratoire d'Astrophysique de Marseille (CNRS/OAMP),
Traverse du Siphon, BP 8, 13376 Marseille Cedex 12 \and
Observatoire de la C\^ote d'Azur, BP 229, 06304 Nice Cedex 4 \and
NBIfAFG, Copenhagen University Observatory, Juliane Maries Vej
30, 2100 Copenhagen \O , Denmark }

   \date{Received October 30th, 2000; Accepted }

   \abstract{
   The T\'elescope \`a Action Rapide pour les Objets Transitoires
   (TAROT-1) has as prime objective the observation of the prompt
   and delayed emission of cosmic gamma-ray bursts (GRBs). We have
   performed a search for optical emission  from 6 GRBs detected by
   BATSE. The positioning error circle was fully covered
   within typically thirty
   minutes after the trigger. No detection of the early
   afterglow phase was made, and magnitude  limits
   in the range of $ \mathrm{m}_{\mathrm{R}} = 13-15 $
   were estimated using 20s exposures.
   These limits are compared
   to optical afterglow data obtained in later phases
    and the results are interpreted in terms of source distances.
   They correspond to
   a median redshift of z = 0.5.
    With HETE-2 and the planned instrument upgrade, TAROT-1 will
    be able to detect the early optical emission of GRBs up to a
    redshift of the order of 5.
}
 \maketitle
%
%________________________________________________________________

\section{Introduction}

Since their discovery more than thirty years ago, Gamma-Ray
Bursts (GRBs) have been intriguing objects  for  theoreticians and
challenging sources for  observers. Their cosmological nature was
firmly established in 1997 by observations performed by the
BeppoSAX satellite (Costa, \cite{Costa99}), which enabled the
first discovery of a faint optical transient associated to the
event (van Paradijs et al. \cite{Vanpara97}). With the possible
exception of GRB 980425, whose association with SN 1998bw remains
in doubt (Pian et al. \cite{Pian99}), the measured GRB source
redshifts are close to, or larger than unity. The events may be
decomposed, in general, in two main phases: a prompt bursting
phase, with no preferred temporal pattern, with a duration
ranging from few milliseconds to several minutes (Paciesas et al.
\cite{Pacie99}); a delayed emission, the afterglow, that in a
first approximation decays according to a power-law of index
$\approx$ -1. BeppoSax wide-field observations were of
fundamental importance for the discovery and the follow-up of the
afterglow phase at different wavelengths (Costa et al.
\cite{Costa97}, van Paradijs et al. \cite{Vanpara97}). The ROTSE
experiment has given evidence for emission at optical frequencies
from the prompt phase, detecting for the first time an optical
burst occurring during the gamma-ray emission of GRB 990123
(Akerlof et al. \cite{Akerl00}).

The detection of X-ray and optical afterglows of cosmic gamma-ray
bursts supports the fireball model (Rees and M\'esz\'aros
\cite{Rees92}, M\'esz\'aros and Rees \cite{Mesz97}, Panaitescu et
al. \cite{Pana98}) as a standard tool to interpret those
observations. In this framework the afterglow emission is
described as synchrotron  and inverse Compton emission of high
energy electrons accelerated during the shock of an
ultra-relativistic shell with the external medium, while the
prompt emission is due to the internal shocks produced by shells
at different speeds within the relativistic blast wave (see Piran
\cite{Piran99} for a review). Both the prompt radiation and early
afterglow phases provide us critical information to establish the
physical processes during the burst itself, as well as the
physical conditions of the surrounding environment (Kumar and
Panaitescu, \cite{Kum00}, Kumar and Piran \cite{Kum2000b}). There
is a general consensus that the fireball plasma is constituted by
$e^-e^+$ pairs and $\gamma$-photons, however the ultimate energy
reservoir and the mechanism of pair creation are still a
challenge to theoretical models.

\begin{table}
\caption[]{Main technical characteristics of TAROT-1}
\label{TAROT1}
\[
         \begin{array}{ll}
            \hline
            \noalign{\smallskip}
\mathrm{Aperture} & 25\,\mathrm{cm} \\
\mathrm{Field\:of\:view} & 1 \degr \times 1 \degr \\
\mathrm{Optical\:resolution} & 20\,\mu\mathrm{m} \\
\mathrm{Mount\:type} & \mathrm{equatorial} \\
\mathrm{Axis\:speed\:} (\alpha\: and\: \delta ) &
                    \mathrm{adjustable,\:up\:to\:}80\degr / \mathrm{s} \\
\mathrm{CCD\:type} & Apogee\:AM13\:\mathrm{with}\:Kodak\:KAF\,1300 \\
\mathrm{CCD\:size} & 1080 \times 1280\,\mathrm{pixels} \\
\mathrm{Pixel\:size} & 15\,\mu\mathrm{m} \\
\mathrm{CCD\:readout\:noise} & \approx 100\,\mathrm{e}^- \\
\mathrm{Readout\:time} & 30\,\mathrm{s} \\
\mathrm{Filter\:wheel} & 6\:\mathrm{pos.:\,Clear,\,
V,\,R,\,I,\,B}+\mathrm{V,\,R}+\mathrm{I}^\mathrm{a} \\
 \noalign{\smallskip}
            \hline
         \end{array}
      \]
\begin{list}{}{}
\item[$^{\mathrm{a}}$] Filters B+V and R+I are broad band filters
covering the spectral range of respectively the Cousin B and V,
and R and I filters.
\end{list}
\end{table}

Since 1991 the Burst and Transient Experiment (BATSE, Fishman
\cite{Fish89}) on board of the Compton Gamma-Ray Observatory
(CGRO) has been detecting about one GRB source per day in the 50
- 300 keV energy band, with fluences ranging from
$10^{-7}\mbox{erg.cm}^{-2}$ to $10^{-3}\mbox{erg.cm}^{-2}$.The
GRB Coordinate Network (GCN, Barthelmy \cite{Bar97}) uses the raw
real time data of BATSE to compute very rapidly an approximate
position of the sources. In spite of the large error circle of the
initial coordinates, $5\degr$ or more, the information on the
source position is transmitted in a very short time, about 4
seconds after the burst triggering by the BATSE detectors. Wide
field telescopes equipped with rapid optical detectors may
quickly point and monitor the source position, searching for an
optical transient, like the detection of GRB 990123 by the ROTSE
experiment (Akerlof et al. \cite{Akerl99}).

The aim of the {\it T\'{e}lescope \`a Action Rapide pour les
Objets Transitoires} (Rapid Action Telescope for Transient
Objects, hereafter TAROT-1, Bo\"er et al. \cite{Boer99}) is the
rapid response to a triggered $\gamma$-ray event, searching for
the detection of the prompt optical counterpart as well as of the
early phases of the afterglow. Although this experiment was
primarily designed to work in conjunction with the HETE satellite
(Ricker et al. \cite{Ricker01}), its wide field permits the
observation of BATSE/GCN alerts using multiple pointings. In this
paper we present the limits obtained for the early afterglow
phases of sources observed during the period 1999 - 2000 detected
by BATSE. These limits are interpreted in light of data obtained
on GRB afterglows and in the framework of the fireball model.

\section{The instrument}

TAROT-1 is a fully autonomous 25 cm aperture telescope. Its
$2\degr$ field of view matches well the HETE uncertainty in
localization of the sources (particularly for the HETE-1 WXM,
since TAROT-1 was designed before its unsuccessful launch). Table
1 summarizes the present main technical characteristics of
TAROT-1.

\begin{figure}
\resizebox{\hsize}{!}{\includegraphics{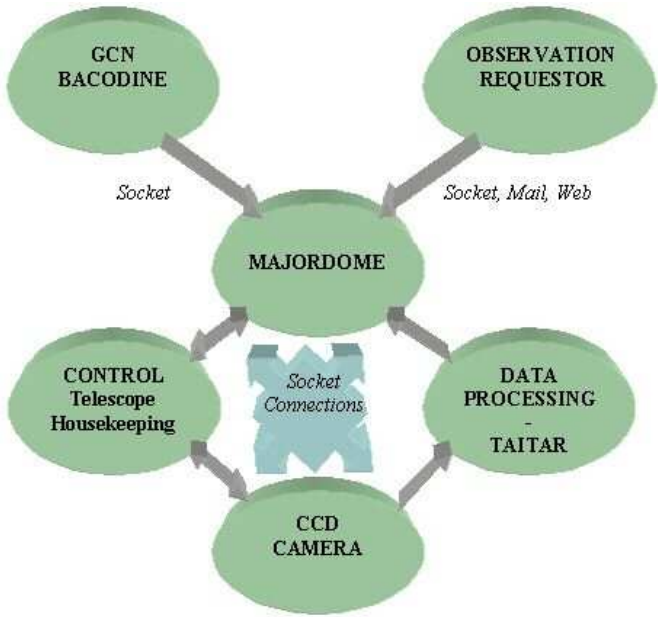}} \caption{The
software modules of TAROT-1 and their interactions from Bringer
et al. \cite{Brin00}. In the case of GRBs, requests are sent to
the MAJORDOME by the GCN (BACODINE). The observations are
immediately scheduled, and processed on line by the TAITAR
software.} \label{SoftTAROT}
\end{figure}

TAROT-1 received about one observable alert from BATSE every ten
days. This is in agreement with the BATSE statistics of GRB event
detection, the relative night$-$day duration and the accessible
area of the sky ($10\degr$ above the North horizon to $20\degr$
for the South part of the sky). Weather conditions prevent
observation during 30\% of the time. Observations are performed
even during the full Moon periods, although with a reduced
sensitivity. TAROT-1 is fully autonomous i.e. there is
practically no human intervention. Figure 1 summarizes the
different software modules and their connections. All  are
functionally independent and the communication is made through
TCP/IP socket processes.

\begin{table*}
\caption[]{TAROT-1 observation log for the GRB source
observations made with CGRO-BATSE} \label{ObsLog}
 \[
 \begin{array}{llllll}
 \noalign{\smallskip}
 \smallskip
 Source\:name&BATSE\:trigger&Error\:circle
 &Matching\:with\:mosaic?&Sky\:conditions&Moon\:phase\\
   &time\:(UT)&\:radius(\,\degr)^\mathrm{a}&(see\:text)& &(\%)\\
 \noalign{\smallskip}
 \hline
 \noalign{\smallskip}
\object{GRB\,990802}& 01:04:44 & 2.1 & Full & Clear& 70\\
\object{GRB\,990807}& 21:55:46 & 2.5 & No\:match & Clear& No\:Moon\\
\object{GRB\,990903}& 19:14:56 & 2.7 & Partial & Partly\:cloudy & No\:Moon\\
\object{GRB\,990915}& 23:15:23 & 2.2 & Full & Clear & No\:Moon\\
\object{GRB\,990917}& 20:12:28 & 2.6 & Partial & Clear & 50\\
\object{GRB\,991002}& 22:49:03 & 2.8 & Full & Clear & 45\\
\object{GRB\,991115}& 18:48:08 & 2.2 & Partial & Cloudy & 45\\
\object{GRB\,991129}& 22:38:08 & 4.0 & No\:match & Clear & 50\\
\object{GRB\,991211}& 04:34:42 & 6.1 & No\:match & Clear & No\:Moon\\
\object{GRB\,000107}& 00:44:09 & 6.8 & No\:match & Clear & No\:Moon\\
\object{GRB\,000111}& 02:41:58 & 5.2 & Partial & Clear & No\:Moon\\
\object{GRB\,000201}& 03:02:09 & 2.1 & Full & Clear & 25\\
\object{GRB\,000221}& 23:39:10 & 2.0 & No\:match& Partly\:cloudy & 85\\
\object{GRB\,000225}& 01:35:29 & 2.3  & Full & Partly\:cloudy & 62\\
\object{GRB\,000226}& 03:51:49 & 2.5 & Partial & Clear & 55\\
\object{GRB\,000229}& 02:45:02 & 2.4 & Full & Cirrus\:clouds? & 35\\
\object{GRB\,000302}& 02:50:25 & 2.1 & No\:match & Clear & No\:Moon\\
\object{GRB\,000312}& 20:50:50 & 2.5 & Full & Clear & 45\\
\object{GRB\,000321}& 21:53:08 & 4.1 & Partial & Partly\:cloudy & 90\\
\object{GRB\,000408}& 02:35:48 & 2.5 & Partial & Clear & No\:Moon\\
\object{GRB\,000418}& 21:30:17 & 5.4 & No\:match & Clear & 100\\
\noalign{\smallskip} \hline
\end{array}
\]
\begin{list}{}{}
\item[$^{\mathrm{a}}$] 68\% confidence radius from the last available BATSE data,
taking into account both statistical and systematic errors.
\end{list}
\end{table*}

The Telescope Control System (TCS) takes care of the instrument,
performs housekeeping, weather monitoring, etc. During the night,
the control is handed over by the MAJORDOME (Bringer et al.
\cite{Brin00}), a software package in charge of the observations.
In the absence of any alert from the GCN, the instrument is in
{\it routine} mode, and observations requested by various
observers are sequenced. Upon receipt of an observable GCN alert
from BATSE, the ongoing observation is interrupted, and the
MAJORDOME sends the telescope and camera parameters to the TCS.
The reconfiguration and pointing of the telescope to its new
position lasts typically 1.5 second. Taking into account the
various delays involved by the GCN (4-6s), and the INTERNET
socket transmission (typically 0.25s), the total reaction time,
from the time of the BATSE trigger to the beginning of data
acquisition, is of about of 7-10s. Since the positioning error
circle of BATSE is larger than the telescope field of view,
observations are performed inside a field of $25^{\sq}\degr$,
covered by a set of $5 \times 5$ mosaic. In the best cases this
total field covers only the 68\% confidence radius of the error
circle, including statistical errors, as discussed in the BATSE
4B catalog (Paciesas et al. \cite{Pacie99}). TAROT-1 reacts to
BATSE\_LOCBURST notices (see the GCN web site:
http://gcn.gsfc.nasa.gov for a description of the various notice
types). After usually 20s to 10min, a new alert is sent by the
GCN, called BATSE\_FINAL, with improved coordinates, producing a
second positioning of TAROT-1, and the start of a new mosaic. If
the difference between the BATSE\_LOCBURST and BATSE\_FINAL
positions is less than $1\degr$ the telescope keeps its current
position. Depending on the burst brightness, the Huntsville BATSE
team may compute a better position, which is usually transmitted
by the GCN 30 to 60min later. TAROT-1 moves accordingly and
eventually reduces the area explored, depending on the actual
uncertainty. During the remaining of the night, a continuous
coverage of the area is maintained, trying to detect and to keep
track of an eventual optical transient.

As soon as an image is obtained, the image processing software
(called TAITAR in figure 1) produces a source catalog. Presently,
the generated list of sources is compared to the USNO catalog
both for astrometric and photometric reductions. Once TAROT-1 has
completed a full mosaic, an e-mail message is sent, and a
scientist on duty is paged to compare the images and to look for
possible new sources. The images are acquired without filter
(clear position in the filter wheel) and with an exposure time of
20s. With the APOGEE camera the readout time was 30s, hence the
total time needed to observe the full mosaic was 21 min.

\section{Observations and reduction}

TAROT-1 observations of GRBs started routinely in 1999, although
the data for the first half of July 1999 has been lost because of
a hardware failure of the media recorder . All observations are
summarized in table 2. The two first columns give the GRB date
and time. The third column give the error circle taken from the
final data reduction provided by the BATSE team, including the
systematic error as given in Paciesas et al. (\cite{Pacie99}).
For several sources, it appears that neither the BATSE\_FINAL nor
the BATSE\_LOCBURST positions were accurate enough to match fully
our mosaic. This indication is given in column 4, while the sky
conditions are depicted in the next column,  with the Moon phase
in the 6th column. The influence of the Moon depends on its phase
and distance to the source, and has an impact on the upper limits
we computed. A phase of 30\% begin to produce a higher background
whatever the Moon position over the sky. No event observed by
TAROT-1 as a response to a BATSE/GCN alert was detected by
another instrument, whatever the wavelength. In six cases, the
68\% error circle has been adequately sampled and the sky
conditions were correct, hence the probability that TAROT-1 never
observed a burst source (i.e. the six sources are outside their
68\% error circle) is 0.0011, assuming a uniform probability
distribution over the error circle. Of course, the detection
probability depends on the unknown source magnitude. Hence the
source detection probability will be $(1-0.68p)^6$, where p
represents the fraction of bursts above the TAROT-1 detection
threshold at the time of observation.

Since no optical sources were detected, a $3\sigma$ upper limit
to the magnitude was calculated by comparing several stars on the
images with the USNO A2.0 catalog. Since observations were
performed without any filter, a correction factor was estimated
in order to provide a R-magnitude upper limit, assuming an energy
spectral index equal to 1. Under this condition, one finds that
the correction to be applied to the measured TAROT magnitude
$\mathrm{m}_{\mathrm{T}}$ is $\mathrm{R} =
\mathrm{m}_{\mathrm{T}} - 0.4$. The limits obtained are listed on
table 3 as a function of the elapsed time since the BATSE
trigger. The delays mentioned in the second column corresponds to
the end of the scan matching the final BATSE source position.

\begin{table}
\caption[]{$3\sigma$ upper limits (transformed in R magnitude) to
the GRB source counterparts for the 6 well observed events
mentioned in table 2.} \label{Limits}
 \[
 \begin{array}{lll}
 \noalign{\smallskip}
 \smallskip
 Source\:name&Time\:since\:trigger&3\sigma\:upper\:limit\\
 & (days)&(R\:magnitude)\\
 \noalign{\smallskip}
 \hline
 \noalign{\smallskip}
\object{GRB\,990802}& 0.0200 & 12.7 \\
                    & 0.0402 & 13.2 \\
\object{GRB\,990915}& 0.0404 & 13.9 \\
                    & 0.0719 & 14.1 \\
\object{GRB\,991002}& 0.0231 & 13.0 \\
                    & 0.0441 & 12.9 \\
                    & 0.0643 & 14.3 \\
                    & 0.0971 & 14.2 \\
                    & 0.1299 & 12.8 \\
\object{GRB\,000201}& 0.0177 & 13.3\\
                    & 0.0317 & 15.3\\
                    & 0.0504 & 15.3 \\
                    & 0.0723 & 15.1 \\
                    & 0.0885 & 14.9 \\
\object{GRB\,000229}& 0.0192 & 14.2 \\
                    & 0.0942 & 14.3 \\
                    & 0.1321 & 15.2 \\
\object{GRB\,000312}& 0.1033 & 14.7 \\
\noalign{\smallskip} \hline
         \end{array}
\]
\end{table}

\section{Discussion}

\begin{table}
\caption[]{The GRB source sample used for comparison of the known
afterglow light curves with our upper limits. Column 3 indicates
the decay index after correction from  } \label{dataused}
\[
         \begin{array}{lll}
            \hline
            \noalign{\smallskip}
           Source&Redshift&References\\
            \noalign{\smallskip}
            \hline
GRB\,970228 & 0.695 & 1, 2, 3, 4 \\
GRB\,970508 & 0.835 & 5 - 10\\
GRB\,971214 & 3.42 & 11 \\
GRB\,980329 & - & 12 \\
GRB\,980519 & - &  13 - 15 \\
GRB\,980613 & 1.096 & 16 - 18 \\
GRB\,980703 & 0.966 & 19 \\
GRB\,990123 & 1.61 & 20, 21 \\
GRB\,990510 & 1.62  & 22, 23 \\
GRB\,991208 & - & 24 - 26 \\
GRB\,991216 & 1.02 & 27 - 33 \\
GRB\,000301c & - & 34 - 40 \\
GRB\,000418 & 1.12 & 41 - 43 \\
 \noalign{\smallskip}
            \hline
         \end{array}
      \]
\begin{list}{}{}
\item[1] Guarnieri et al., \cite{Guar97}
\item[2] Metzger et al. \cite{Metz97}
\item[3] Pedichini et al. \cite{Pedi97}
\item[4] van Paradijs et al. \cite{Vanpara97}
\item[5] Fruchter et al. \cite{Fruch97}
\item[6] Galama et al. \cite{Gal97}
\item[7] Garcia et al. \cite{Gar97}
\item[8] Djorgovsky et al. \cite{Djo97}
\item[9] Mignoli et al. \cite{Mign97}
\item[10] Pedersen et al. \cite{Peder98a}
\item[11]Diercks et al. \cite{Diercks98}
\item[12]Reichart et al. \cite{Rei99}
\item[13]Bloom et al. \cite{Bloom98a}
\item[14]Gal et al. \cite{Gal98}
\item[15]Vrba et al. \cite{Vrba00}
\item[16] Djorgovsky et al. \cite{Djo98}
\item[17] Halpern et al. \cite{Halp98}
\item[18] Hjorth et al. \cite{Hjort98}
\item[19] Bloom et al. \cite{Bloom98b}
\item[20] Akerlof et al. \cite{Akerl99}
\item[21] Galama et al. \cite{Gal99}
\item[22] Harrison et al. \cite{Hari99}
\item[23] Staneck et al. \cite{Sta99}
\item[24] Kuulkers et al. \cite{kuu00}
\item[24] Castro-Tirado et al. \cite{Castro99}
\item[25] Jensen et al. \cite{Jensen99a}
\item[26] Masetti et al. \cite{Mas99}
\item[27] Djorgovsky et al. \cite{Djo99}
\item[28] Dolan et al. \cite{Dolan99}
\item[29] Garnavich et al. \cite{Garna99}
\item[30] Henden et al. \cite{Henden99}
\item[31] Jensen et al. \cite{Jensen99b}
\item[32] Jha et al. \cite{Jha99}
\item[33] Sagar et al. \cite{Sagar00}
\item[34] Bernabei et al. \cite{Berna00}
\item[35] Fynbo et al. \cite{Fyn00}
\item[36] Gal-Yam et al. \cite{Galy00}
\item[37] Garnavich et al. \cite{Garna00}
\item[38] Halpern et al. \cite{Halp00}
\item[39] Mohan et al. \cite{Mohan00}
\item[40] Veillet and Bo\"er \cite{Veillet00}
\item[41] Henden \cite{Henden00}
\item[42] Henden and Klose \cite{HendKlo00}
\item[43] Mirabal et al. \cite{Mirabal00}

\end{list}
   \end{table}

TAROT-1 upper limits on the optical flux of GBR 990123, the only
source until now detected both during its prompt and afterglow
phase, were compared with ROTSE data. Figure 2 displays TAROT-1
upper limits and ROTSE measurements (Akerlof et al.
\cite{Akerl99}), as well as later data (Galama et al.
\cite{Gal99}).

\begin{figure}
\resizebox{\hsize}{!}{\includegraphics{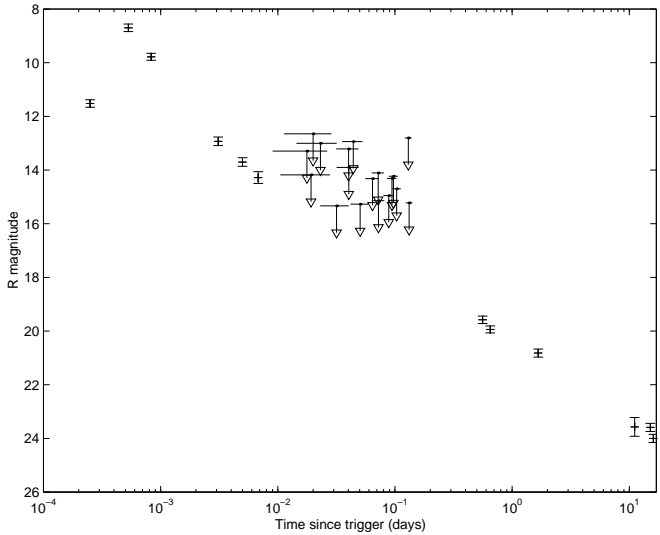}}
\caption{Comparison between the TAROT upper limits (arrows) and
the light curve of GRB 990123 (error bars). Both the prompt and
the afterglow emission are shown. Data from Akerlof et al.
(\cite{Akerl99}) and from Galama et al. (\cite{Gal99})}
\label{GRB990123}
\end{figure}

Another possibility is to compare TAROT-1 upper limits with data
obtained for GRB afterglows at later times. The result of this
comparison is displayed in figure 3, whereas the basic data and
references are given in table 4. We note that in the time region
observed by TAROT-1, \object{GRB\,970508}, \object{GRB\,980519},
\object{GRB\,991208}, and \object{GRB\,991216}, in addition to
\object{GRB\,990123},would have been detected by TAROT, if we
extrapolate the late epoch afterglow to early times. Of course,
this is a crude extrapolation which does not take into account
possible breaks in the light curves.
\begin{figure}
\resizebox{\hsize}{!}{\includegraphics{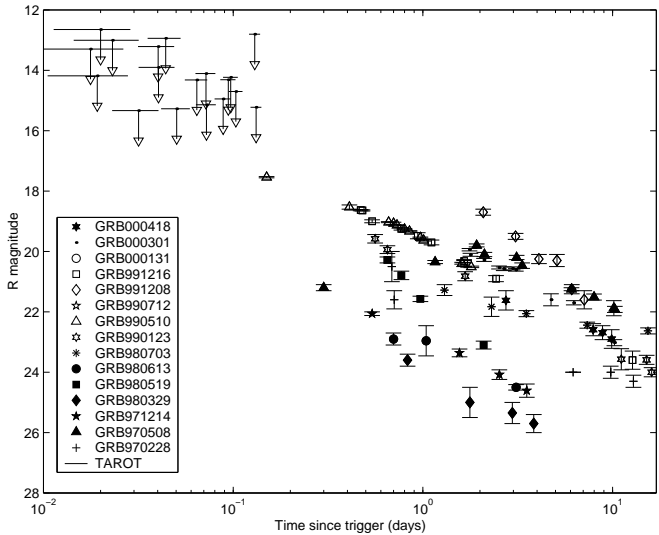}}
\caption{Comparison between the TAROT upper limits (arrows, all
data) and the afterglow light curve of GRB sources detected at
optical wavelengths (table 4). In this figure, the light curves
have not been corrected} \label{limits}
\end{figure}

We tried to evaluate the possible effect of the source distance
on the burst afterglow detection by TAROT-1. To perform this task
we have to correct for the galactic absorption and for the source
distance (Bo\"er and Gendre \cite{Boer00}). We have scaled the
flux of sources with known distances to a common redshift of z =
1,  applying corrections for distance, time dilation and
absorption in our galaxy (Schlegel et al. \cite{Schl98}). We
assume a flat cosmology with ${\Omega}_{\Lambda} = 0.7$ and
$\mathrm{H}_0 = 65 \mathrm{km.s}^{-1}\mathrm{.Mpc}^{-1}$. Table 5
gives the corrections computed and the decay indexes. The
corrected light curves are displayed in figure 4 together with
 TAROT-1 upper limits. As it can be seen, extrapolating to the
proper time domain, TAROT-1 would have detected 5 events, if their
redshift would have been equal to z = 1.

These data can also be used to estimate the limiting TAROT-1
distance of detection. Assuming a limiting magnitude of 15 one
hour after the burst, the corresponding distance is $\mathrm{z} =
1.2$ for the brightest burst and 0.1 for the faintest one, the
median redshift being 0.5.

\section{Conclusions and perspectives}

In this paper, the lower R-magnitude limit for the early
afterglow light curves of GRB is computed, for sources observed
by TAROT-1 after its first year of routine operation. These
limits are compared with the optical afterglow data obtained so
far. A maximum source distance of $\mathrm{z} = 1.2$ was estimated
for observations made with TAROT-1 one hour after the GRB main
event begins. The accuracy of HETE-2 (Ricker et al.,
\cite{Ricker01}) or SWIFT (Gehrels, \cite{Gehrels00}) will allow
to start the observation within few seconds after the GRB triggers
the detectors, while for BATSE we had to scan the error circle, an
operation lasting 30 minutes. Given the expected HETE-2 detection
rate of 40 events per year in the anti-solar direction, TAROT-1
may expect to observe a source location once every month. In
order to estimate the maximum source distance TAROT-1 can reach,
we extrapolate the light curves plotted on figure 4 to a time of
five minutes after the burst onset. In this case, the median
redshift corresponding to a limiting magnitude of 15 is
$\mathrm{z} = 1.2$.

\begin{figure}
\resizebox{\hsize}{!}{\includegraphics{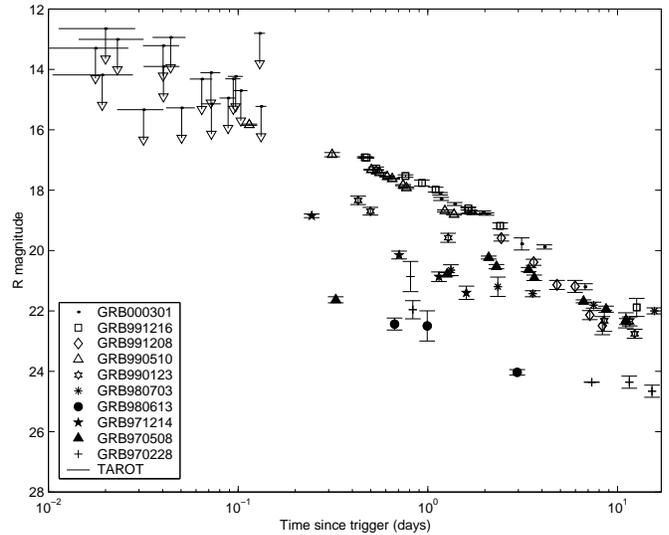}}
\caption{Comparison between the TAROT upper limits (arrows, all
data) and the afterglow light curve of GRB sources. The light
curves have been normalized to a common distance corresponding to
a redshift of 1, according to the method described in the text
(Bo\"er and Gendre \cite{Boer00}) and using the data from table
5. We took into account the K-correction, distance and time
dilation, and absorption from our Galaxy (table 5).}
\label{limitscorrected}
\end{figure}

\begin{table*}
\caption[]{Distance and galactic absorption correction factors for
the GRB sources of known distance. These quantities are used to
normalize the GRB light curves to a common distance corresponding
to a redshift of 1 (see text). The corrected power law decay
index is tabulated column 5 (for details see Bo\"er and Gendre,
\cite{Boer00})} \label{Optics}
 \[
 \begin{array}{lllll}
 \noalign{\smallskip}
 \smallskip
            &\multicolumn{3}{c}{Correction\:factors}\\
            Source&Correction\:for&K-&Galactic& Corrected\:light\:curve\\
                   &distance       &Correction          &absorption&decay\:index\\
 \noalign{\smallskip}
 \hline
 \noalign{\smallskip}
\object{GRB\,970228}& 0.41 & 1.0& 0.61& 1.10\pm 0.11 \\
\object{GRB\,970508}& 0.64 & 1.0& 0.05& 1.17\pm 0.21 \\
\object{GRB\,971214}& 20.2& 1.2& 0.06& 1.20\pm 0.27 \\
\object{GRB\,980613}& 1.10 & 1.0& 0.22& 1.00\pm 0.01 \\
\object{GRB\,980703}& 0.92& 1.0& 0.72& 1.17\pm 0.25 \\
\object{GRB\,990123}& 3.23 & 0.9& 0.04& 1.44\pm 0.07 \\
\object{GRB\,990510}& 3.28 & 0.9& 0.53& 1.54\pm 0.15 \\
\object{GRB\,991216}& 1.05 & 1.0& 1.67 & 1.22\pm 0.04 \\
\noalign{\smallskip}
            \hline
         \end{array}
      \]
\end{table*}

In addition,  the detector will soon be replaced by a camera
based on a Thomson 7899 CCD. It will have a total noise of
$10\mathrm{e}^-$, for a readout time of 1 second, allowing to
reach the 17th R magnitude in 10 seconds. With this new device
TAROT-1 will be able to monitor the GRB optical light curves, not
only during the prompt event, but also the transition between the
prompt and afterglow phase. After five minutes, even in the worst
case discussed in section 4 ($ \mathrm{z} = 0.1$, one hour after
the burst onset with the previous camera) the TAROT-1 distance
limit will correspond to a redshift of 2, and the median redshift
of 5.6, while after 1 hour, this distance will still correspond
to a median redshift of 2.2 (0.5 with the previous camera).

Already in its current configuration TAROT-1 is able to give
stringent limits to the early afterglow phase. In conjunction
with a satellite like HETE-2 or SWIFT, the transition between the
prompt and delayed mode will be monitored, allowing to acquire
critical data on GRB sources.

\begin{acknowledgements}

We would like to give tribute to our collaborator G\'erard
Calvet, engineer at the INSU/DT, who tragically died during a
mission at the Plateau de Bure IRAM observatory. The {\it
T\'{e}lescope \`a Action Rapide pour les Objets Transitoires}
(TAROT-1) has been founded by the {\it Centre National de la
Recherche Scientifique} (CNRS), {\it Institut National des
Sciences de l'Univers} (INSU) and the Carlsberg Fundation. It has
been built with the support of the {\it Division Technique} of
INSU (INSU/DT). We thank the anonymous referee for his helpful
comments.

\end{acknowledgements}

\end{document}